\documentclass[11pt]{article}
\usepackage{moriond,epsfig}
\usepackage{amsmath,amssymb}

\makeatletter
\def\address#1{\expandafter\def\expandafter\@aabuffer\expandafter
{\@aabuffer \small\it #1\relax 
\begin{figure}[h]
\begin{center}
\psfig{file=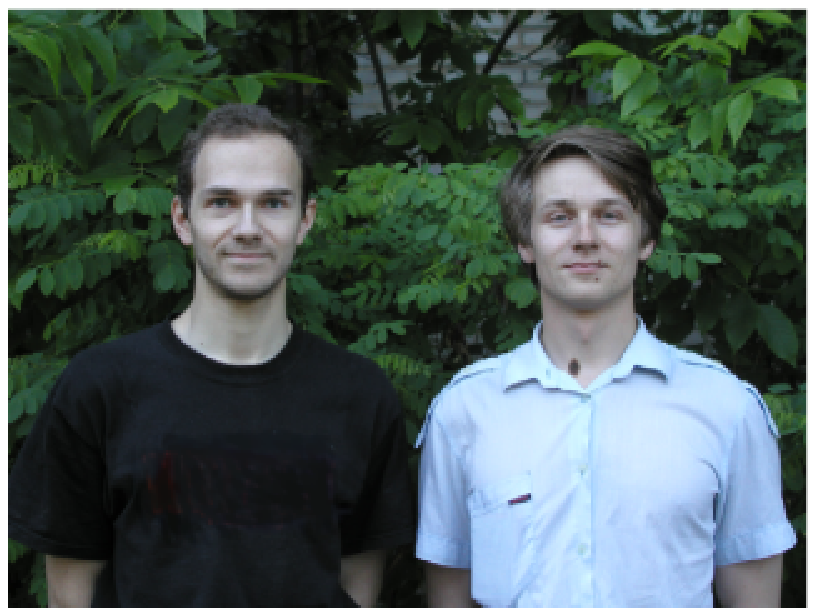,width=2in}
\end{center}
\end{figure}
\par\vspace{1em}}}
\makeatother

\newcommand{\Real}{\mathop\mathrm{Re}\nolimits}
\newcommand{\Imag}{\mathop\mathrm{Im}\nolimits}

\newcommand{\sm}[1]{{\scriptscriptstyle \rm #1}}

\newcommand{\la}{\langle}
\newcommand{\ra}{\rangle}

\begin{document}
\vspace*{4cm}
\title{T-ODD CORRELATIONS IN $\pi\to e\nu_e\gamma$ DECAY}

\author{ F.~L.~Bezrukov, D.~S.~Gorbunov }

\address{    Institute for Nuclear Research of the Russian Academy of Sciences,\\
    60th October Anniversary prospect 7a, Moscow 117312, Russia}

\maketitle\abstracts{ The transverse lepton polarization asymmetry in
  $\pi_{e2\gamma}$ decay may probe T-violating interactions beyond the
  Standard Model.  Dalitz plot distributions of the expected effects
  are presented and compared to the contribution from the Standard
  Model final state interactions.  We give an example of a
  phenomenologically viable model, where a considerable contribution
  to the transverse lepton polarization asymmetry arises.}

\section{Introduction}

T-violation beyond the Standard Model is usually searched for in
decays forbidden by time reversal symmetry.  Another way to probe
T-violation is measurement of T-odd observables in allowed decays of
mesons.
Widely considered T-odd observables are transverse muon polarizations
($P_T$) in $K\to\pi\mu\nu$ and $K\to\mu\nu\gamma$ decays.  There is no
tree level SM contribution to $P_T$ in these decays, so they are of a
special interest for search for new physics.  Unfortunately, $P_T$ is
not exactly zero in these decays even in T-invariant theory~---
electromagnetic loop corrections contribute to $P_T$ and should be
considered as a background.  There are no experimental evidence for
nonzero $P_T$ in these processes at present time, but the sensitivity of the
experiments has not yet reached the level of SM loop contributions.

In this talk we discuss the decay $\pi\to e\nu\gamma$.  Within the
Standard Model, T-violation in this process does not appear at tree
level, but interactions, contributing to it, emerge in various
extensions of SM.  We shall demonstrate that $\pi_{e2\gamma}$ decay is
an attractive probe of new physics beyond the Standard Model.
Depending on the model, $\pi_{e2\gamma}$ decay may be even more
attractive than usually considered $K_{l2\gamma}$ decays.  The similar
decay $\pi_{\mu2\gamma}$ is analized in the article
\cite{Bezrukov:2002zc}, where one can also find the detailed
bibliography.

Though $\pi_{e2\gamma}$ decay has very small 
branching ratio (it is of order $10^{-7}$), we find that 
the distribution of the transverse electron polarization over the
Dalitz plot significantly overlaps with the distribution of
differential branching ratio, as opposed to the situation with
$K_{\mu2\gamma}$ decay.  Moreover, the contribution of FSI
(final-state interactions related to SM one-loop diagrams) to the
observable asymmetry, being at the level of $10^{-3}$, becomes even
smaller in that region of the Dalitz plot, where the contribution from
new effective T-violating interaction is maximal. Thus,
$\pi_{e2\gamma}$ decay is potentially quite interesting probe of
T-violation. 

To demonstrate that pion decays may be relevant processes where the
signal of new physics may be searched for, we present a simple model
of heavy pseudoscalar particle exchange leading in the low energy
limit to the T-violating four fermion interaction.  We find the
constraints on the parameters of this model coming from various other
experiments and describe regions of the parameter space which result
in large T-violating effects in $\pi_{l2\gamma}$ decays.  Depending on
the parameters of the model, an experiment measuring transverse lepton
polarization with pion statistics of $10^5\div10^{10}$ pions for
$\pi_{\mu2\gamma}$ decay and $10^8\div10^{13}$ pions for
$\pi_{e2\gamma}$ decay is needed to detect the T-violating effects
(taking into account statistical uncertainty only and assuming ideal
experimental efficiencies).


\section{T-violating effect in $\pi^+_{e2\gamma}$ decay}
\label{sec:2}

\begin{figure}[t]
  \begin{center}\begin{tabular}{ccc}
    \psfig{figure=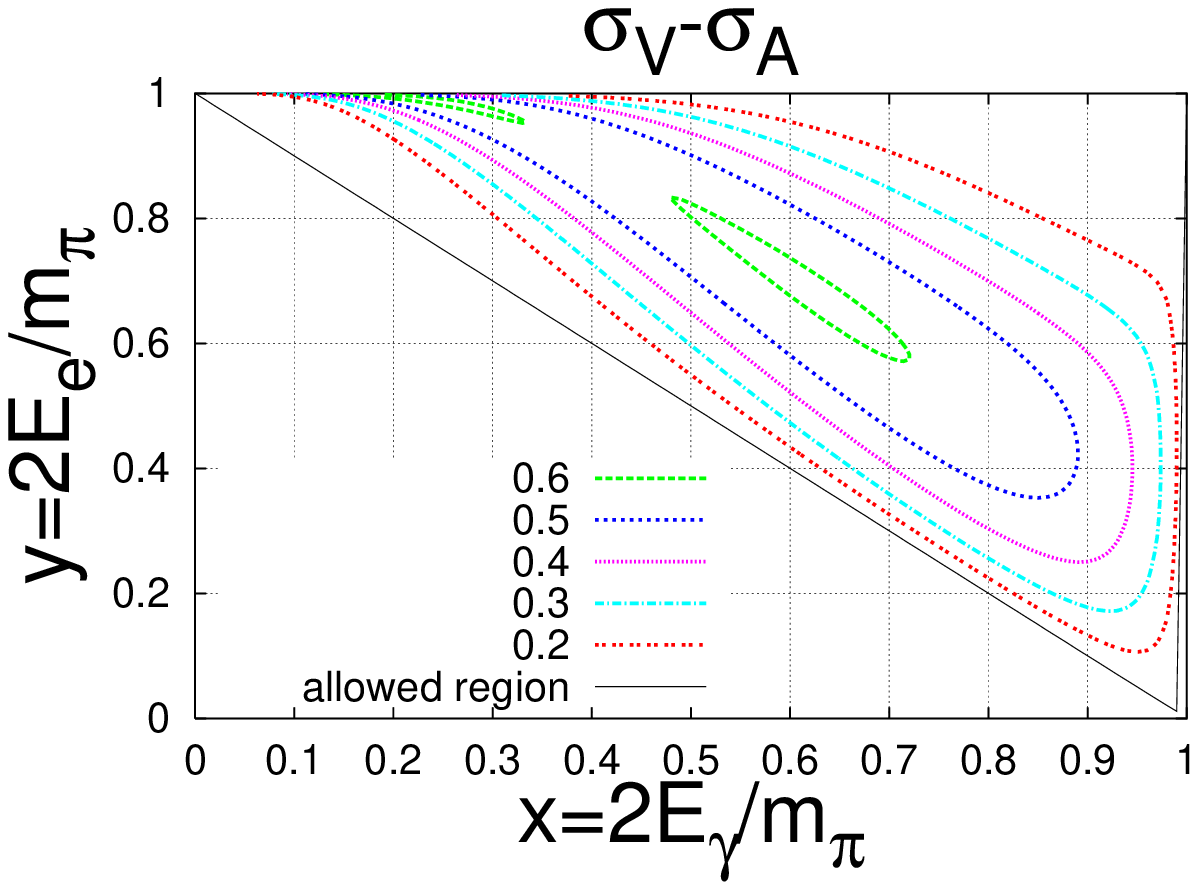,width=0.3\textwidth} &
    \psfig{figure=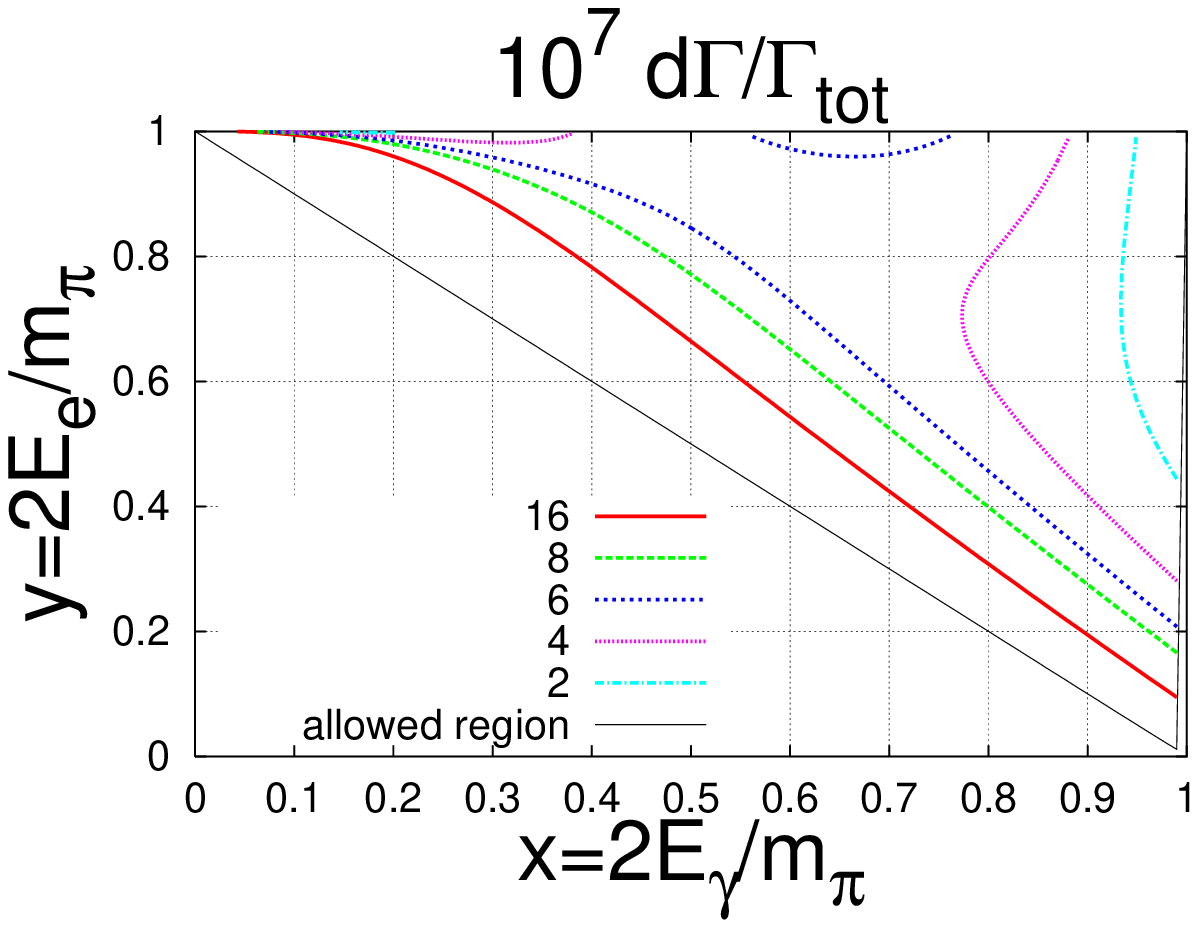,width=0.3\textwidth} &
    \psfig{figure=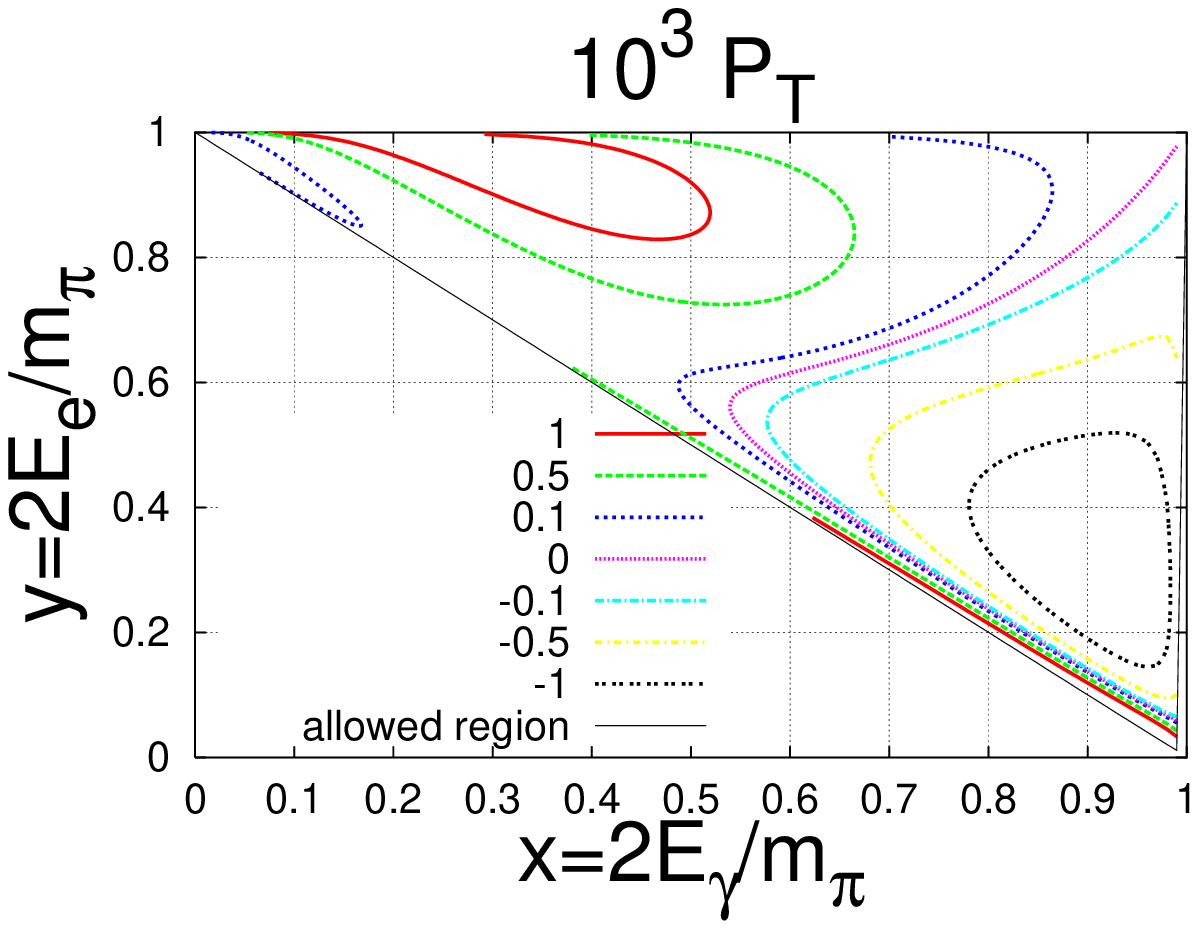,width=0.3\textwidth} \\
    (a) & (b) & (c)
  \end{tabular}\end{center}
\end{figure}

Let us consider the simplest effective four-fermion
interaction
\begin{equation}\label{4-fermion}
  {\cal L}_{\mathrm{eff}} =
  G^e_{\sm P}\bar{d}\gamma_5u\cdot\bar{\nu}_e(1+\gamma_5)e
  +\mathrm{h.c}.
\end{equation}
that may be responsible for the T-violating effects in pion physics
beyond the Standard Model.  Indeed, the imaginary part of the constant
$G^e_{\sm P}$ contributes to transverse lepton polarization.  

For the transverse electron polarization asymmetry (polarization in the
direction $\vec{e}_T = {\vec{q}\times \vec{p}_l}/|\vec{q}\times \vec{p}_l|$)
\begin{gather*}
  P_T(x,y)={d\Gamma(\vec{e}_T)-d\Gamma(-\vec{e}_T) \over 
  d\Gamma(\vec{e}_T)+d\Gamma(-\vec{e}_T)}
  =[\sigma_V(x,y)-\sigma_A(x,y)]\cdot\Imag[\Delta^l_P]\\
   \Delta^e_P = \frac{\sqrt{2}G^e_P}{G_F\cos\theta_c}\cdot
     \frac{B_0}{m_l}\;,\qquad B_0=-{2\over (f^0_\pi)^2}\la 0|\bar{q}q|0\ra=
   {m_\pi^2\over m_u+m_d}\approx2\mathrm{GeV}\;.  
\end{gather*}

The contour-plot of $[\sigma_V-\sigma_A]$ as a function of $x$ and $y$
is presented in the Figure (a).

As one can see, in a large region of kinematic variables,
$[\sigma_V-\sigma_A]$ is about $0.5$.  This means that transverse
electron polarization $P_T$ for this process, is of the same order as
$\Imag[\Delta_P^e]\simeq5\times10^3\cdot\Imag[G_P^e/G_F]$.  It is
worth noting that the region of the Dalitz plot where large
T-violating effect might be observed, significantly overlaps with the
region where the partial decay width $\Gamma(\pi_{e2\gamma})$ is
saturated (cf.\ Figures~(a) and (b)).  This is in contrast to the
situation with T-violation in $K_{\mu2\gamma}$ decay, where the
analogous overlap is small, so the differential branching ratio in the
relevant region is smaller than on average.

Theoretical background to the observation of $P_T$ caused by
interaction of the type \eqref{4-fermion} appeares from the
contribution from final-state interactions (FSI) --- one-loop diagrams
with virtual photons.  The value of this contribution in the Standard
Model is presented in Figure~(b).

For the $\pi_{e2\gamma}$ decay, the ($x,y$)-distributions of FSI
contribution and the contribution from the four-fermionic
interaction~(\ref{4-fermion}) differ in shape.  Specifically, part of
the region with maximal $P_T$ from four-fermion interaction
\eqref{4-fermion} corresponds to the region of small $P_T$ from FSI\@.
This implies that if measured, $P_T$ distribution could probe
T-violating interaction~(\ref{4-fermion}) with an accuracy higher than
$\Imag[\Delta^e_P]\sim 10^{-3}$
($\Imag[G_P^e/G_F]\sim2\times10^{-7}$).  Again, this is not the case
for $K_{\mu2\gamma}$ decay.

\subsection{Constraint from $\pi\to l\nu$ decays}
\label{sec:2.2}

The interaction term~(\ref{4-fermion}) not only gives rise to
T-violation in $\pi\to e\nu_e\gamma$ decay but also contributes to the
rate of $\pi\to l\nu_l$ decays.  Since the ratio of leptonic decays of
the pion has been accurately measured, the coupling constants $G_P^e$
is strongly constrained.  This constraint can be evaded by introducing
a similar coupling to muon and muon neutrino with the corresponding
constant $G_P^\mu = \frac{m_\mu}{m_e}G_P^e$.

Note that to the leading order in $\Delta_P$, only the real parts of
the coupling constants $G_P^\mu$ and $G_P^e$ are constrained, while
constraints on imaginary parts are weaker.  Thus, for general
$G_P^\mu$ and $G_P^e$ (if the mentioned hierarchy does not hold) one
obtains $|\Real[\Delta_P]|\lesssim10^{-3}$ and
$|\Imag[\Delta_P]|\lesssim0.03$.  Hence in this case experiments aimed
at searching for T-violation in $\pi_{e2\gamma}$ decays should have
sufficiently large statistics: the total number of charged pions
should be $N_\pi\gtrsim10^{11}$.

In models with $\Real[G_P]\sim\Imag[G_P]$ and without $\mu-e$
hierarchy, the bound from $\pi\to l\nu$ decays implies
$|\Imag[\Delta_P]|\lesssim10^{-3}$, which significantly constrains
possible contribution of the new interaction~(\ref{4-fermion}) to
T-odd correlation in $\pi_{e2\gamma}$ decay.  Namely, the contribution
to the $\pi_{e2\gamma}$ decay should be of the same order or weaker
than one from the Standard Model FSI\@.  Nevertheless, as we
discussed, in the case of $\pi_{e2\gamma}$ the difference in ($x,y$)
distributions of FSI and four-fermion contributions may allow one to
discriminate between the two if they are of the same order of
magnitude, and even if the contribution of four-fermion
interaction~(\ref{4-fermion}) is somewhat weaker.  To test
four-fermion interaction~(\ref{4-fermion}) at the level allowed by
$\pi\to e\nu_e$, i.e., at the level of $10^{-3}$, one has to collect
not less than $10^{13}$ charged pions, assuming statistical
uncertainty only.

Overall, in the case of the hierarchy muon and electron couplings,
decays $\pi\to l\nu$ do not constrain new T-violating interactions
which can be searched for in relatively low statistics experiments,
$N_\pi\gtrsim10^8$ for $\pi_{e2\gamma}$ and $N_\pi\gtrsim10^5$ for
$\pi_{\mu2\gamma}$.  In the worst case of no hierarchy and $\Real
G_P\sim\Imag G_P$, new T-violating interactions have little chance to
be observed, and need high statistics experiments.

\section*{Acknowledgments}
The authors are indebted to V.~N.~Bolotov, Yu.~G.~Kudenko and
V.~A.~Rubakov for stimulating discussions.  The work is supported in
part by CPG and SSLSS grant 00-15-96626, by RFBR grant 02-02-17398 and
by the program SCOPES of the Swiss National Science Foundation,
project No.~7SUPJ062239.  The work of F.B.\ is supported in part
by CRDF grant RP1-2364-MO-02.  The work of D.G.\ is supported in
part by the RFBR grant 01-02-16710 and INTAS YSF 2001/2-142.

\section*{References}

\end{document}